# Magnetic anomalies in single crystalline $ErPd_2Si_2$

E.V. Sampathkumaran, Niharika Mohapatra, Kartik K. Iyer, C.D. Cao[1,2], W. Löser[1], and G. Behr[1]

Tata Institute of Fundamental Research, Homi Bhabha Road, Colaba, Mumbai – 400005, India.
[1] Leibnitz-Institut für Festkörper-und Werkstoffforschung Dresden, Postfach 270116, D-01171 Dresden, Germany
[2] Department of Applied Physics, Northwestern Polytechnical University, Xi'an 710072, P.R. China

## Abstract

Considering certain interesting features in the previously reported [166]Er Mössbauer effect and neutron diffraction data on the polycrystalline form of $ErPd_2Si_2$, crystallizing in $ThCr_2Si_2$-type tetragonal structure, we have carried out magnetic measurements (1.8 to 300 K) on the single crystalline form of this compound. We observe significant anisotropy in the absolute values of magnetization (indicating that the easy axis is c-axis) as well as in the features due to magnetic ordering in the plot of magnetic susceptibility ($\chi$) versus temperature (T) at low temperatures. The $\chi(T)$ data reveal that there is a pseudo-low dimensional magnetic order setting in at 4.8 K, with a three-dimensional antiferromagnetic ordering setting in at a lower temperature (3.8 K). A new finding in the $\chi(T)$ data is that, for H//<110> but not for H//<001>, there is a broad shoulder in the range 8-20 K, indicative of the existence of magnetic correlations above 5 K as well, which could be related to the previously reported slow-relaxation-dominated Mössbauer spectra. Interestingly, the temperature coefficient of electrical resistivity is found to be isotropic; no feature due to magnetic ordering could be detected in the electrical resistivity data at low temperatures, which is attributed to magnetic Brillioun-zone boundary gap effects. The results reveal complex nature of the magnetism of this compound.

The ternary rare-earth (R) compounds, particularly containing Ce, Eu and Yb, derived from $ThCr_2Si_2$-type tetragonal structure [1] are of great interest for the past few decades. Among these, the series $RPd_2Si_2$ [2] attracted considerable interest in the literature due to interesting phenomena discovered [3,4,5] among these compounds like pressure-induced superconductivity and anomalous valence transitions. The so-called 'normal' R ions have not been studied in depth, as 4f-orbitals in these are well-localised and hence unusual electron correlation effects are not expected. In this article, we focus our attention on one such member, namely $ErPd_2Si_2$. Following original neutron diffraction work by Bazela et al [6], careful magnetization (M), [166]Er Mössbauer effect and neutron diffraction measurements were performed [7] on the polycrystalline form of this compound. While magnetic susceptibility ($\chi$) and neutron diffraction data reveal that this compound orders antiferromagnetically (AF) below 4.8 K in a complex fashion with the moments on Er aligning along c-axis, the magnetic hyperfine split pattern persists in the Mössbauer spectra even at 16 K which was interpreted in terms of fluctuation of Er moments with a long relaxation time. In addition, the temperature dependence of magnetic reflections in neutron diffraction data as evidenced by the growth of diffusion peaks suggests that, interestingly, a pseudo-bidimensional order takes place at 4.8 K, and that the magnetic ordering becomes three-dimensional at a lower temperature. In order to throw more light on the magnetism of this compound, it is essential to carry out magnetic studies on single crystals, which is the motivation of the present studies. The results obtained are quite revealing.

The $ErPd_2Si_2$ single crystal was grown by a floating zone technique with radiation heating from a polycrystalline feed rod prepared by arc melting stoichiometric amounts of Er (>99.98%), Pd (>99.95%) and Si (99.9999%). A coarse grained $ErPd_2Si_2$ seed was utilized. The floating-zone crystal growth process with radiation heating was performed in a laboratory type apparatus URN-2ZM (MPEI, Moscow) with a vertical double ellipsoid optical configuration and a 5 kW air-cooled xenon lamp positioned at the focal point of the lower mirror. The growth process proceeded in a chamber under flowing Ar. Axially symmetric counter-rotation of crystal (33 rpm) and feed rod (20 rpm) and a growth velocity of 10 mm/h were employed to grow the single crystal of about 6 mm in diameter and 40 mm in length. The orientation of single crystals was determined by the X-ray Laue back-scattering method. Microstructure and crystal perfection were examined by optical metallography and scanning electron microscopy (SEM). Energy Dispersive X-ray (EDX) analysis confirmed that the composition of the single crystal is very close to ideal stoichiometry. The specimens with the two rod axes, <001> and <110>, employed in the present studies, were cut from the crystal prepared. Magnetic susceptibility studies in the presence of a magnetic field (H) of 5 kOe were carried out in the temperature (T) interval 1.8 - 300 K employing a commercial vibrating sample magnetometer (Oxford Instruments, UK). The same magnetometer was employed for isothermal magnetization studies at selected temperatures. In addition, electrical resistivity ($\rho$) measurements were carried out (1.8 - 300 K) (also in the presence of magnetic fields) with the help of a physical property measurements system (Quantum Design, USA) by a conventional four probe method employing silver paint for making electrical contacts of the leads with the sample.

In figure 1, we show the plot of inverse $\chi(T)$ for the two rod orientations, H//<001> and H//<110>. The plots are linear at high temperatures (> about 25 K) and the deviation from linearity below 25 K is more prominent for the latter orientation. Since the crystal field splitting falls in the range 30 - 100 K [7], one would expect that this factor plays a role for this deviation. Possibly, short range magnetic order also plays a role as inferred from isothermal M data (see below). The effective magnetic moment ($\mu_{eff}$) obtained from the high temperature linear region is 9.4 and 9.85 $\mu_B$ respectively, very close to the free ion value. Corresponding values of paramagnetic Curie temperature are 5 and - 17 K respectively; the sign suggests dominance of ferromagnetic interaction along c-axis and AF interaction along the basal plane.

The low temperature data (Fig. 2) are quite revealing. For H//<001>, there is a peak at 4.8 K as in polycrystals, characteristic of AF ordering. This peak is broad as seen in figure 2. However, for H//<110>, this broad peak is absent near 5 K, but instead a *much sharper* peak is seen at 3.8 K. This downward shift in ordering temperature for this direction supports the idea [7] that magnetic interaction setting at 4.8 K is pseudo-bi-dimensional. There are two more interesting features in figure 2: (i) For H//<001>, there is another noticeable drop in $\chi$(T) at 2.5 K as the temperature is lowered, as though there is another subtle change in magnetic structure at this temperature. This manifests itself as a weak upturn in $\chi$(T) for H//<110>. (ii) For H//<110> alone, a broad peak in the $\chi$(T)-plot appears in the range 8 -20 K, which was not observed in the polycrystals in the earlier literature [7]. Since the higher-lying Kramers doublets are above 30 K (Ref. 7) from the ground state, we believe that this peak is not due to crystal-field effects, but magnetic in origin. We therefore propose that there are magnetic moment fluctuations with a finite antiferromagnetic component in the basal plane. This may be corroborated with the magnetic hyperfine splitting observed in the Mössbauer spectra around the same temperature range. Finally, as the temperature is lowered, the values for H//<001> gets larger than that for the other orientation, suggestive of anisotropy in the magnetic behavior.

In order to throw more light on the magnetic behavior, we show the isothermal M behavior up to 120 kOe in figure 3 at various temperatures. There are qualitative differences in the shapes of M(H) curves for the two orientations establishing anisotropic magnetic behavior. (i) For H parallel to easy axis (figure 3a), at 1.8 K, M saturates for H > 30 kOe, whereas for the orientation along the basal plane (figure 3b), the variation of M with H is more gradual without any evidence for saturation in the measured field range. The value of the magnetic moments at 120 kOe for the latter orientation is less than that for the former (7.6 $\mu_B$). These values are less than the theoretical value (9 $\mu_B$) implying that the ground state is not fully degenerate due to crystal-field effects. (ii) In the data at 1.8 K, there is a metamagnetic transition around 5 kOe for H//<001>, whereas this transition is shifted to the high-field region (70-100 kOe) for H//<110>. Other noteworthy findings in M(H) data are: (i) The presence of metamagnetic feature for H//<001> is consistent with a net antiferromagnetism [7] along the easy axis; (ii) the M shows interestingly a weak hysteresis at 1.8 K in the range 3-7 kOe for H//<001>, whereas, the weak hysteresis is observable only at high fields for H//<110> (see figure 3 insets); the origin of hysteresis at intermediate fields is not clear. (iii) The high-field saturation of M persists above 5 K as well (see 6 and 10 K data in figure 3a) for the former orientation without metamagnetic transition feature, as though it is easy to induce ferromagnetism (with zero coercive field) by the application of a magnetic field. (iv) For the latter orientation, the essential 1.8 K-features (weak spin-reorientation and weak hysteresis at high fields) could be observable even at 10 K, a finding that implies that the short-range magnetic correlations/fluctuations persist well above 5 K – an important conclusion made in the previous paragraph.

We have also measured electrical resistivity as a function of temperature for the two orientations, I//<001> and I//<110> (where I is the excitation current). The absolute values of $\rho$ for the former orientation are found to be three times smaller than the latter orientation.. However, when normalized to respective 300 K values, the plots for the two orientations lie one over the other, as shown in figure 4. This implies that the determination of the exact spacing of the voltage leads could be erroneous due to spread of silver paint (employed for making electrical contacts). The main point of emphasis is that no feature due to magnetic ordering could be observable at low temperatures. The resistivity $\rho$ varies rather smoothly across magnetic transitions without any drop. It therefore appears that there is a formation of magnetic Brillouin-zone gaps due to the complexity of the  modulated AF structure [7], the contribution of which apparently offsets the decrease due to the loss of spin-disorder contribution. It may also be remarked that there is no change either in the values of $\rho$ or in the features of $\rho(T)$ with the application of magnetic fields up to 140 kOe. This establishes the robustness of possible magnetic gap in this system.

Summarizing, we have presented the results of magnetization studies on the single crystals of $ErPd_2Si_2$. While the magnetic behavior is anisotropic, the scattering process interestingly is isotropic. The magnetization data reveal that the magnetic ordering setting in near 4.8 K is not isotropic and that there is an additional (possibly three-dimensional) magnetic transition at 3.8 K. There appears to be another magnetic anomaly near 2.5 K. In addition, the data provide evidence for additional short range magnetic correlations in the range 4.8 – 20 K, which could also account for Mössbauer spectral features reported in the literature. There is a weakly hysteretic metamagnetic transition at 1.8 K appearing at low fields for H//<001>, which is shifted to very high field range for H//<110>. Thus, the results reveal complex nature of the magnetism of this compound. Possibly, there is an interesting interplay between interlayer and intralayer magnetic couplings and crystal-field effects in this compound, which makes this compound attractive for further studies.

**Acknowledgment:** Two of us (W.L. and G.B.) express their gratitude to financial support by SFB 463 of the Deutsche Forschungsgemeinschaft  and one of us (C.D.C) to the Alexander von Humboldt-stiftung.

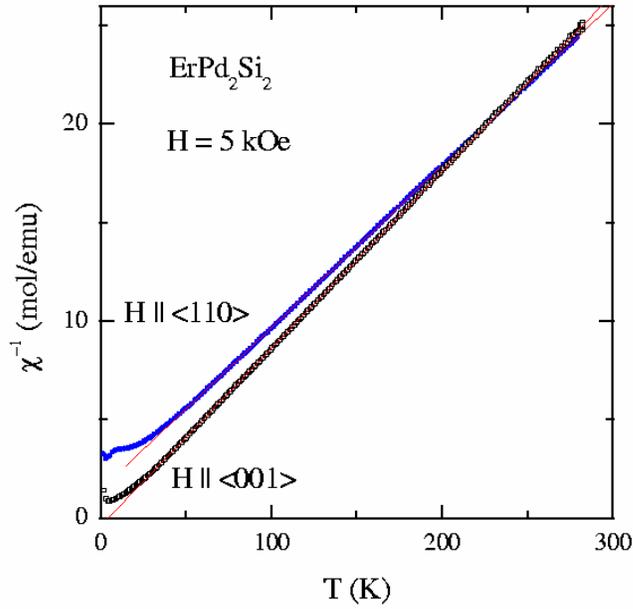

Figure 1:    Inverse susceptibility as a function of temperature for two orientations of single crystalline ErPd₂Si₂, measured in the presence of a magnetic field of 5 kOe. A line is drawn through the high temperature linear region.

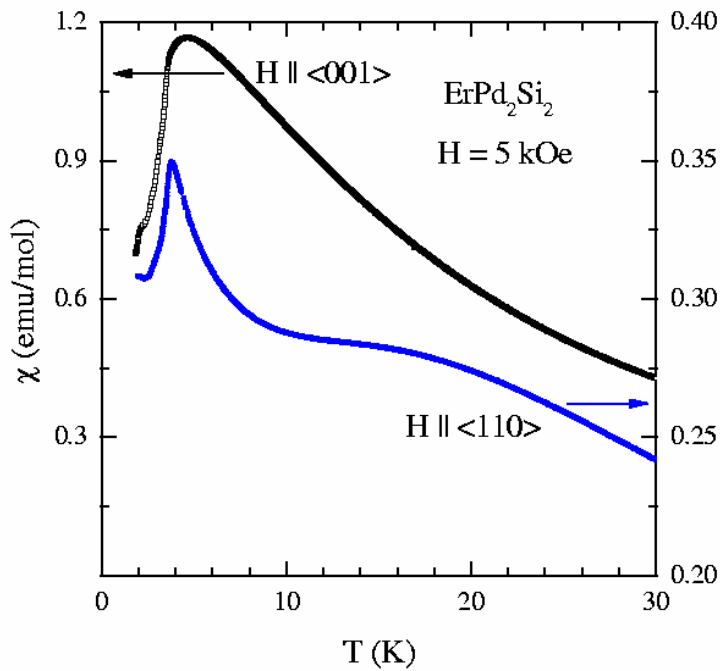

Figure 2:    Magnetic susceptibility as a function of temperature (below 30 K) for two orientations of single crystalline ErPd₂Si₂, measured in the presence of a magnetic field of 5 kOe.

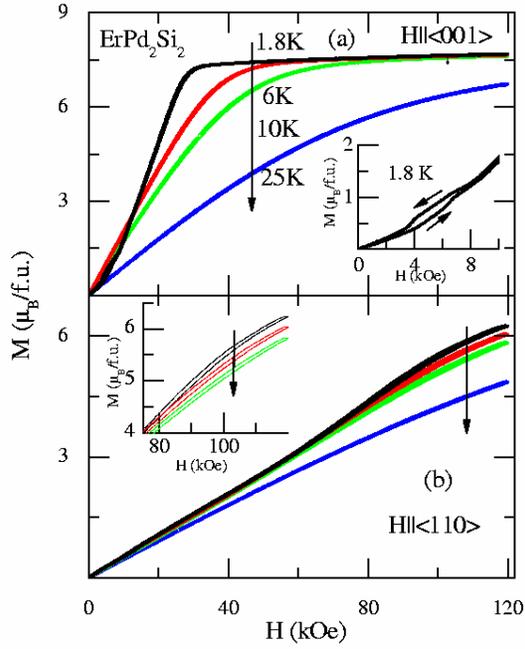

Figure 3: Isothermal magnetization behavior at selected temperatures for two orientations of single crystalline ErPd$_2$Si$_2$. The curves for up and down field variations overlap, except a weak hysteresis in the range 3 – 7 kOe for 1.8 K alone for H//<001> (see figure **a**, inset), and in the high-field region for T ≤ 10K for H//<110> (see figure **b** inset). The downward vertical arrows are shown to identify the curves with increasing temperature.

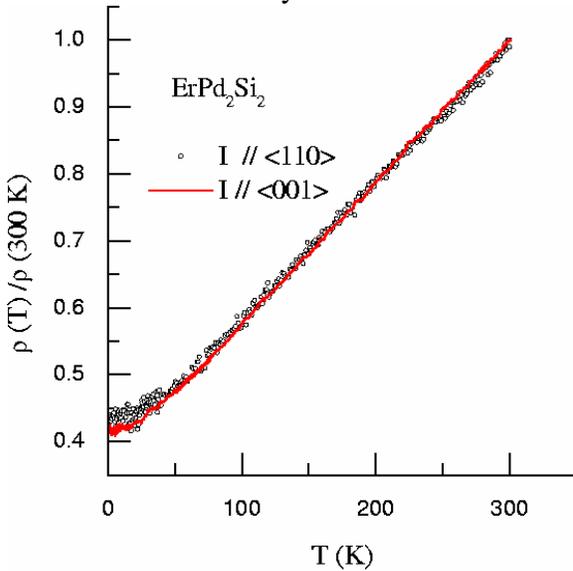

Figure 4: Electrical resistivity as a function of temperature for two rod axes, (I//<001> and I//<110> (where I is the excitation current)), for ErPd$_2$Si$_2$ single crystals. The values are normalized to respective 300 K values.